\documentclass[epj]{svjour}

\usepackage{psfig}
\usepackage{graphics,epsfig}
\input epsf
\usepackage{amssymb}

\newcommand{\be}{\begin{equation}}
\newcommand{\ee}{\end{equation}}
\newcommand{\bea}{\begin{eqnarray}}
\newcommand{\eea}{\end{eqnarray}}

\newcommand{\dis}{\displaystyle}

\begin{document}
\title{\hfill {\small FZJ--IKP(TH)--2004--19, HISKP-TH-04-22}\\ 
\\ 
Flatt{\'e}-like distributions and the $a_0(980)/f_0(980)$ mesons}


\author{V. Baru\inst{^1},  J. Haidenbauer\inst{^2}, 
C. Hanhart\inst{^2}, A. Kudryavtsev\inst{^1}, Ulf-G. Mei{\ss}ner\inst{^{2,3}}}

\institute{
Institute of Theoretical and Experimental Physics,
117259, B. Cheremushkinskaya 25, Moscow, Russia 
\and 
Institut f\"ur Kernphysik (Theorie), Forschungszentrum J\"ulich,
D-52425 J\"ulich, Germany
\and
Helmholtz-Institut f\"ur Strahlen- und Kernphysik (Theorie),
Universit\"at Bonn, Nu\ss allee 14-16, D-53115 Bonn, Germany
}

\date{Received: date / Revised version: date}

\abstract{
We explore the features of Flatt{\'e}-like parametrizations. 
In particular, we demonstrate that the 
large variation in the absolute values of the coupling constants to the
$\pi\eta $ (or $\pi\pi$) and $K\bar K$ channels for 
the $a_0$(980) and $f_0$(980) mesons that one can find in the
literature can be explained by  a specific scaling behaviour of the 
Flatt{\'e} amplitude for energies near the $K\bar K$ threshold. We 
argue that the ratio of the coupling constants can be much better 
determined from a fit to experimental data.  
}

\maketitle

\section{Introduction}

In spite of a long history of investigations since their discovery the nature 
of the light scalar mesons $f_0(980)$ and $a_0(980)$ is far from being 
understood \cite{Buggrep,Klempt,Torn,Oller,Bev,Ani}. 
Usually to get information about the properties of these   
resonances the so-called Flatt{\'e} parametrization \cite{Flatte}
for the differential mass distributions is used. This parametrization was 
first introduced by  Flatt{\'e}
for the description of the $\pi\eta$ invariant mass distribution 
near the $K\bar K$ threshold where the $a_0(980)$ scalar-isovector resonance 
is located. 

The Flatt{\'e} differential mass distribution is a slightly modified
relativistic version of the Breit-Wigner distribution. It reads, e.g.,
for the $a_0$ channel 
\be
\frac{d\sigma_i}{dm} \propto \left|\frac{m_R\sqrt{\Gamma_{\pi\eta}\Gamma_i}}
{m_R^2-m^2-im_R(\Gamma_{\pi\eta}+\Gamma_{K\bar K})}\right|^2,
\label{Flatte}
\ee
with the partial widths $\Gamma_{\pi\eta}=\bar g_{\eta}q_{\eta}$ and 
\bea
\Gamma_{K\bar K}=\bar g_K \sqrt{m^2/4-m^2_K} 
\nonumber
\eea
above threshold and 
\bea
\Gamma_{K\bar K}=i\bar g_K\sqrt{m_K^2-m^2/4} 
\nonumber
\eea
below threshold,
respectively. The subscript $i$ in Eq. (\ref{Flatte})
labels the $\pi\eta$ and/or $K\bar K$ channels. Furthermore, 
$m_R$ is the nominal mass of the resonance, $m$ is the
invariant mass ($m^2=s$) and $q_\eta$ is the corresponding center-of-mass
momentum in the ${\pi\eta}$ system. 
$\bar g_\eta$ and $\bar g_K$ are dimensionless coupling 
constants that are related to the dimensional coupling constants $g_{\pi\eta}$ 
and $g_{K\bar K}$ commonly used in the literature by 
$\bar g_\eta=g^2_{\pi\eta}/(8\pi m^2_R)$ and 
$\bar g_K=g^2_{K\bar K}/(8\pi m^2_R)$, respectively.

Since the position of the $a_0$ peak is located far from the $\pi\eta$ threshold and 
the peak is relatively narrow, we may consider the inelastic width $\Gamma_{\pi\eta}$ to
be approximately constant for energies near the $K\bar K$ threshold. The width 
$\Gamma_{K\bar K}$, on the other hand, varies rapidly near this threshold.
The shape
of the Flatt{\'e} distribution (\ref{Flatte}) is determined by three free 
parameters $m_R$, $\Gamma_{\pi\eta}$ (or $\bar g_{\eta}$) and $\bar g_K$,
which should be fixed from a fit of Eq. (\ref{Flatte}) to the
experimental differential mass distribution.
As was stressed already in Ref. \cite{Flatte}, if the value of $\bar g_K$ is reasonably 
large, then the $\Gamma_{K\bar K}$ term in the denominator of Eq. (\ref{Flatte}) 
will suppress the cross section for masses below the 
$K\bar K$ threshold, thus narrowing the $\pi\eta$ mass distribution. It was also
mentioned in Ref. \cite{Flatte} that ``the differences between the shapes with 
$\Gamma_{\pi\eta}$ = 80 MeV and $\Gamma_{\pi\eta}$ = 300 MeV are relatively slight''.

More than two decades have passed since the paper by  Flatt{\'e} and a 
wealth of different experimental data concerning the $f_0(980)$ and $a_0(980)$ 
resonances has been obtained in the meantime. 
Nevertheless the uncertainties in the
parameters extracted for both resonances remain large!
To demonstrate this we compile in Tables 1 and 2 some results of analyses 
for the $a_0$ \cite{Teige,Bugg,Amsler,SNDa,KLOEa,AchKi}
and $f_0$ \cite{SND,CMD2,KLOE,E791,WA102,OPAL} 
mesons where Flatt{\'e} or Flatt\'e-like parametrizations
like the one by Achasov (see, e.g. Refs. \cite{AchKi,AchGu} and references 
therein) are utilized. Note that these different
parameterizations are equivalent in the nonrelativistic limit,
i.e. near the $K\bar K$ threshold, which means that they all lead
to the same analytical form and the parameters of a particular 
distribution can be re-expressed by those of the other distribution. 
We show this explicitly for Achasov's distribution in Appendix. 

Table 1 clearly demonstrates, for the $a_0$ meson,
that the resulting absolute values of
$\bar g_\eta$ ($\Gamma_{\pi\eta}$) and $\bar g_K$ differ significantly 
for different analyses. At the same time it also reveals that the ratios 
of the coupling constants, 
$R=\bar g_K/\bar g_{\eta}=g_{K\bar K}^2/g_{\pi\eta}^2$,
are more or less consistent with each other  
for practically all the parameter sets extracted from the  experimental data. 
For the $f_0(980)$ meson the situation is very
similar for most of the results shown in Table 2. 
We should mention that in case of the results of Refs. \cite{E791,OPAL} for the
$f_0$ and of Ref.\cite{SNDa} for the $a_0$, which deviate so strongly from the general
trend, the value of $\bar g_K$ is afflicted with  large errors, cf. Ref. \cite{SNDa,E791,OPAL}.

In the present paper we investigate the features of Flatt\'e or Flatt\'e-like
parametrizations. In the course of this 
we demonstrate that the relative stability for the ratio $R$ and the 
large variations in the absolute values of the coupling constants and masses, 
evidenced by the different fits as mentioned above, can be understood. It is 
simply a consequence of a specific scaling behaviour of the Flatt{\'e} amplitude 
for energies near the $K\bar K$ threshold. 
The corresponding scale transformation is introduced in Sect. 2 and we 
discuss its implications for the elastic (i. e. $\pi\pi$ or $\eta\pi$) scattering 
amplitude near the $K\bar K$ threshold 
and for the effective range parameters of the $K\bar K$ channel.  
  
In Sect. 3 we focus on the interplay between resonance structure and 
threshold. In particular, we investigate the movement of the poles as a function of
the coupling strength to the $K\bar K$ channel and we examine the corresponding
results for the ($\pi\pi$ or $\eta\pi$) scattering amplitude and 
phase shifts.
 
Sect. 4 deals with the concrete cases of the $a_0$(980) and $f_0$(980)
mesons.  
Employing various Flatt{\'e} parametrizations for those mesons from the literature 
we exemplify that the resulting phase shifts are indeed very similar 
for most parameter sets, despite of the fact that the resonance parameters themselves 
differ drastically. Thus, the actual results for the $a_0$(980) and $f_0$(980)
clearly reflect the typical scaling behaviour that we derived for such
Flatt{\'e}-type parametrizations. 
The paper ends with some concluding remarks. 

\begin{table}[t]
\begin{center}
\vskip 0.2cm
\begin{tabular}{|c|c|c|c|c|c|c|c|}
\hline
Ref.&$m_R$&$\Gamma_{\pi\eta}$&$\bar g_{\eta}$&$\bar g_{K}$&$R$&$E_{R}$&$\alpha$\\
\hline
\cite{Teige}&1001&70&0.218&0.224&1.03&9.6&0.276\\
\cite{Bugg}&999&143&0.445&0.516&1.16&7.6&0.106\\
\cite{Amsler}&999&69&0.215&0.222&1.03&7.6&0.221\\
\cite{SNDa}$^a$&995& 125 &0.389  &1.414  &3.63  &3.7  &0.027  \\
\cite{KLOEa}$^a$&984.8  &121  &0.376  & 0.412 &1.1 &-6.5  &-0.28   \\  
\cite{AchKi}$^a$&1003&153&0.476&0.834&1.75&11.6&0.096\\
\cite{AchKi}$^a$&992&145.3&0.453&0.56&1.24&0.6&0.006\\
\hline
\end{tabular}
\caption{Flatt\'e parameters for the $a_0$(980) meson taken from the literature. 
The values of $m_R$, $\Gamma_{\pi\eta}$ and $E_{BW}$ are 
given in MeV. 
Values for Refs. labeled  with the superscript $^a$ are based on
Achasov's parametrization \protect\cite{AchKi}, cf. also the Appendix. 
}
\label{Table_1}
\end{center}
\end{table}

\begin{table}[t]
\begin{center}
\vskip 0.2cm
\begin{tabular}{|c|c|c|c|c|c|c|c|}
\hline
Ref.&$m_R$&$\Gamma_{\pi\pi}$&$\bar g_{\pi}$&$\bar g_{K}$&$R$&$E_{R}$&$\alpha$\\ 
\hline
\cite{SND}$^a$&969.8&196&0.417&2.51&6.02&-21.5&-1.35\\
\cite{CMD2}$^a$&975&149&0.317&1.51&4.76&-16.3&-1.00\\
\cite{KLOE}$^a$&973&256&0.538&2.84&5.28&-18.3&-1.07\\
\cite{E791}&977&42.3&0.09&0.02&0.22&-14.3&-0.66\\
\cite{WA102}& -- &90&0.19   &0.40&2.11&--&--\\
\cite{OPAL} &957  &42.3&0.09&0.97&10.78&-34.3&-1.60\\
\hline
\end{tabular}
\caption{Flatt\'e parameters for the $f_0$(980) meson taken from the literature. 
The values of $m_R$, $\Gamma_{\pi\eta}$ and $E_{BW}$ are 
given in MeV. 
Values for Refs. labeled  with the superscript $^a$ are based on
Achasov's parametrization \protect\cite{AchKi}, cf. also the Appendix. 
}
\label{Table_2}
\end{center}
\end{table}

\section{ The Flatt{\'e} distribution and scaling behaviour}
\label{distrib}

Let us concentrate here on energies near the $K\bar K$ threshold.
In this region the resonance part of the elastic scattering amplitude for the
channel with the light particles
($\pi\eta$ for the $a_0$(980) case or $\pi\pi$ for the $f_0$(980) meson)
can  be written in a nonrelativistic form by 
\be
f_{el}=-\frac{1}{2q}\frac{\Gamma_P}{E-E_{BW}+i\frac{\Gamma_P}{2}+i\bar g_K\frac{k}{2}} \ .
\label{nonrel}
\ee
This nonrelativistic expression can be derived starting out both from the
relativistic Flatt{\'e} formula (see Eq.(\ref{Flatte})) and from 
the Flatt{\'e}-like distributions like the one introduced by Achasov
\cite{AchKi} which takes into
account the so called finite width corrections due to $\pi\eta $ (or
$\pi\pi$) and $K\bar K$ loops
(see Appendix for more details). 
Here we use the notation $\Gamma_P=q\bar g_P$ for the inelastic width, 
where $P$ stands for the ${\pi\eta}$ or ${\pi\pi}$ channels and $q$ is the
corresponding center-of-mass momentum.  
%
The energies are defined with respect to the $K\bar K$ threshold, i.e. 
\bea
&E&=\sqrt{s}-2m_K, \ \ m_R=2m_K+E_R, 
\eea
where $\dis m_K=({m_{K^+}+m_{K_0}})/{2}$  and $m_R$ is the nominal mass of the
resonance. 
The parameter $E_{BW}$, which would correspond to the peak position of the standard 
nonrelativistic Breit-Wigner (BW) resonance, i.e. in the limit $\bar g_K\to
0$, is defined as
\bea
\nonumber 
&&E_{BW}=E_R
\eea for the Flatt\'e distribution ($E_{BW}$ for the Achasov
distribution is defined by Eq. (\ref{achparam}) in the Appendix).  
In addition the relative $K\bar K$ momentum, $k$, that also 
appears in Eq. (\ref{nonrel}), is given by  $k=\sqrt{m_K E}$.
Note that $k$ is imaginary for $E < 0$, i.e. for $\sqrt{s}<2m_K$. 

We 
consider here the case of elastic scattering in the resonance approximation 
corresponding to Eq. (\ref{nonrel}), i.e. without any background contributions. 
Then, for $E>0$ the elastic cross section is given by 
\be
\sigma_{el}=4\pi|f_{el}|^2=\frac{\pi \bar g_{P}^2}{(E-E_{BW})^2+
(\Gamma_{P}+\bar g_Kk)^2/4} \ .
\label{epos}
\ee
For $ E<0 $ we get 
\be
\sigma_{el}=\frac{\pi \bar g_{P}^2}{(E-E_{BW}-\kappa\bar g_K/2)^2+\Gamma_{P}^2/4} \ ,
\label{eneg}
\ee
where $\kappa=\sqrt{m_K|E|}$ is the modulus of the imaginary $K\bar K$ momentum $k$.
In the vicinity of the $K\bar K$ threshold
we may omit $E$  
in Eqs. (\ref{epos}) and (\ref{eneg}) 
which leads us to the following 
approximate expressions for the cross section valid near the $K\bar K$ threshold 
\bea
&\sigma_{el}&=\frac{4\pi}{q^2}\frac{1}{\alpha^2+(1+Rk/q)^2}, \ E>0 \ ,   
\nonumber \\
&\sigma_{el}&=\frac{4\pi}{q^2}\frac{1}{(\alpha+R\kappa /q)^2+1}, \ \  E<0 \ ,
\label{scale}
\eea
where $\alpha=2E_{BW}/\Gamma_P$ and $R=\bar g_K/\bar g_P$. 

We see that in the approximation (\ref{scale}) the cross section does not 
depend on all three 
Flatt\'e parameters, $E_{BW}$ and the coupling constants $\bar g_{P}$ and 
$\bar g_K$, but only on the ratios $R$ and $\alpha$. 
Neglecting terms which are of higher order than $k$ ($\kappa$) we get from 
Eqs. (\ref{scale}) 

\be
\sigma_{el}=\frac{4\pi}{q^2}\frac{1}{1+\alpha^2}X,
\label{cones}
\ee
where $$X=1-\frac{2R}{1+\alpha^2}\frac{k}{q} ~~ \rm{for}~~ E>0,$$ 
and $$X=1-\frac{2R\alpha}{1+\alpha^2}\frac{\kappa}{q}~~\rm{for}~~ E<0. $$
Obviously the right and the left slopes (with respect to $k$ ($\kappa$)) 
of the elastic cross section at the $K\bar K$ 
threshold are given only in terms of the ratios $R$ and $\alpha$. 
Note that the sign of the slope
$d\sigma/d\kappa$ for $E<0$ depends on the sign of $E_{BW}$ ($\alpha$), whereas
the slope $d\sigma/dk$ for $E>0$ is always negative. 
Thus, the near-threshold momentum dependence of the invariant mass 
spectrum allows to determine both parameters $R$ and $\alpha$
unambiguously from data. 

The fact that the result in Eqs. (\ref{scale}) depends only on the
ratios $R$ and $\alpha$ means that it 
is invariant with respect to the scale transformation
\be
E_{BW}\to \lambda E_{BW},~~ \Gamma_P\to \lambda \Gamma_P,~~ 
\bar g_K\to \lambda\bar g_K \ .
\label{scatra}
\ee
On the other hand, $\sigma_{el}$ as calculated from the original
Flatt\'e distribution (\ref{nonrel}) is not scale-invariant, 
cf. Eqs. (\ref{epos}) and (\ref{eneg}). Here the corresponding transformed 
elastic scattering amplitude has the form 
\be
f_{el}=-\frac{1}{2q}\frac{\Gamma_P}{(\frac{E}{\lambda})-E_{BW}+i\frac{\Gamma_P}{2}+i\bar g_K\frac{k}{2}}.
\label{rescaled}
\ee
Obviously, this expression reduces to the scale-invariant form in the
limit $|E| \to 0$. 

We should emphasize at this stage that the above considerations hold only if the
resonance is located close to the $K\bar K$ threshold, i.e. within an energy
range where also the above expansion is valid. That seems to be the case for 
basically all the parametrizations as one can see from the values of $E_R$
in Tables 1 and 2. 
If the resonance is located away from the threshold region then there 
will be no scale invariance. But then, of course, one would
not use a Flatt\'e distribution either. 

In order to illustrate the effect of the scale transformation we give, 
in Figs. \ref{Fig1ab}, some examples of differential mass distributions in 
the scaling limit (\ref{scale}) and for different values of the parameter $\lambda$.
The starting parameters $\Gamma_P$, $\bar g_K$ and $E_{BW}$ are those of 
Ref. \cite{CMD2}, given in Table 2.
 Fig. \ref{Fig1ab}a shows the results for negative $E_{BW}$ and 
Fig. \ref{Fig1ab}b those for positive $E_{BW}$. One can see that scaling is
practically fulfilled for the energy interval of roughly $\pm$ 25 MeV
around the $K\bar K$ threshold. Indeed, the cross sections above the threshold
are basically the same for all $\lambda$ up to the highest considered 
excess energy of 100 MeV. For energies below the threshold and $E_{BW} <$ 0
scaling breaks down soon and there are already significant differences in 
the cross sections for energies around $E \approx -50$ MeV, whereas for the
case of $E_{BW} >$ 0 those differences remain small down to even -100
MeV.

Of course, the range of scaling depends to some extent on the
parameters $R$ and $\alpha$. 
In order to demonstrate this we show, in Fig. \ref{Fig1a0},  results obtained with the
parameters of Ref. \cite{Bugg}.
Here the ratio $R$  is more than three times smaller than for the case
considered above. It is clear from  Fig.  \ref{Fig1a0} that the scaling
behaviour is practically limited to the region very near to the $K\bar K$ threshold 
when the ratio $R$ is relatively small.

Since the shape of the mass distribution near the $K\bar K$-threshold is 
scale-invariant, as we just found, 
it is interesting to discuss what happens with the position of the singularities 
of the amplitude (\ref{nonrel}) as a function of the scaling parameter $\lambda$. 
The position of the poles in the complex $k$-plane can be given  
in terms of the $K\bar K$ scattering length $a$ and effective range $r_e$:
\be
k_{1,2}=\frac{i}{r_e}\pm\sqrt{-\frac{1}{r_e^2}+\frac{2}{ar_e}} \ .
\label{poles}
\ee
Those effective range parameters can then be related to the 
Flatt{\'e} parameters $E_{BW}$, $\Gamma_P$ and $\bar g_K$:
\be
a=-\frac{\bar g_K}{2(E_{BW}-i\Gamma_P/2)},~~r_e=-\frac{4}{m\bar g_K}.
\label{scaeff}
\ee

Note that in the Flatt{\'e} parametrization of the scattering amplitude
the effective range is always negative. This is in contradiction to the usual
interpretation of the effective range as interaction range \cite{xxx}, which 
should be positive of course. We should mention, however, that in the case of potential 
models the effective range can be positive as well as negative (see, e.g., 
Ref. \cite{yyy}). A positive effective range can be also obtained if one takes
into account the so-called finite width corrections as it is done in the 
Flatt\'e-like distribution of Achasov, cf. the Appendix and
also the work of Kerbikov \cite{Kerbikov}. 

As is seen from Eqs. (\ref{scaeff}) the scattering length $a$ remains unchanged by 
the scale transformation (\ref{scatra}) whereas the 
effective range $r_e$ rescales according to $r_e\to r_e/\lambda$. 
From Eq. (\ref{poles}) we conclude that in the limit of small $\lambda$ 
both roots are located near the points $k=\pm\sqrt{2/ar_e}$, i.e. practically symmetric 
around the point $k=0$. This corresponds to the case where the 
pure BW resonance dominates the cross section.
In the limit of large $\lambda$ the effective range is getting small and the nearest pole to 
the point $k=0$ is located at $k_1\approx 1/a$. 
The second pole is at $k_2=2i\lambda/r_e$ and plays
practically no role anymore for the physics around the $K\bar K$ threshold. 
In Figs. \ref{Fig2ab} we show the trajectories of the poles of the scattering amplitudes 
(\ref{rescaled}) as a function of the 
parameter $\lambda ~~ (0.1\le\lambda<\infty)$ for the same sets of parameters 
($\bar g_P,\bar g_K, E_{BW}$) used for the results in Figs. \ref{Fig1ab}. 

The limiting, scale-invariant form of the cross section as given by the Eqs. (\ref{scale})
corresponds to the scattering amplitude in the zero range approximation, i.e. to 
\be
f_{el}=\frac{1}{R}~\frac{1}{-a^{-1}-ik} \ .
\label{dlina}
\ee
The scattering length $a$ can be expressed in terms of the ratios $R$ and $\alpha$ 
and the relative momentum of the light particles at the $K\bar K$ threshold, $q_{th}$:
\be
a=-\frac{1}{q_{th}}\frac{R}{\alpha-i} \ .
\label{zereff}
\ee

Thus, since the ratios $R$ and $\alpha$  
can be extracted from a study of the near-threshold momentum dependence of 
the invariant mass spectrum, a determination of the (complex) $K\bar K$ scattering length 
is also feasible.     

\section{Resonance - threshold interplay: poles, cross sections and phase shifts}

In a previous paper \cite{evidence} we discussed the concept of ``pole counting''
suggested by  Morgan \cite{morgan} as a tool for near-threshold resonance classification. 
It was demonstrated that the existence of a pair of poles in the complex $k$-plane in the vicinity 
of the threshold of the heavy particles corresponds to the situation where a resonance has a 
large admixture of an elementary (bare) state made up of a quark-antiquark pair, say. 
In contrary the situation when there
is only one pole near the threshold point $k=0$ corresponds to the molecule-like picture of a resonance.

The
role of the $K\bar K$ threshold for the shape of the differential mass spectra becomes more 
transparent if one studies the pole positions in the $k$-plane as a function of the 
strength of the coupling to the $K\bar K$ channel, $\bar g_K$, while keeping the other two 
parameters, $E_{BW}$ and $\Gamma_P$, fixed. Corresponding results are presented in 
Fig. \ref{Fig3ab}. In parallel we also look at the 
behaviour of the elastic cross section in the channel of the light particles. Those
results are shown in Figs. \ref{Fig4ab}. As an example we take  the Flatt\'e parameters of
Ref. \cite{OPAL}, given in Table 2, as starting point. The coupling strength $\bar g_K$
is varied by multiplying it with a factor $\gamma$ and we label the corresponding curves 
in the figures with the values chosen for $\gamma$. 

In the $k$-plane the position of the poles is given by Eq. ({\ref{poles}). 
Note, that in the limit $\bar g_K\to 0$ we get $a\to 0$ and $r_e\to -\infty$, 
see Eq. (\ref{scaeff}). 
However, combining Eqs. ({\ref{poles}) and (\ref{scaeff}) we see that 
in the limit $\bar g_K\to 0$ the poles of the amplitude (\ref{nonrel}) are at 
$k_{1,2}=\pm \sqrt{2/ar_e}=\pm \sqrt{m_K(E_{BW}-i\Gamma_P/2)}$.
These are simply the poles of the pure BW amplitude. They are located in the 2nd and 4th 
quadrants of the $k$-plane and they are symmetric with respect to the point $k=0$, cf.
Fig. \ref{Fig3ab}. Note that in the limit $\bar g_K\to 0$ the resonance is completely 
decoupled from the $K\bar K$ system. 
The proximity of the resonance state to the $K\bar K$ threshold is then just accidental. 

The elastic cross sections corresponding to the  uncoupled case are shown as 
dashed lines in Fig. \ref{Fig4ab}. 
For $E_{BW}<0$ (Fig. \ref{Fig4ab}a) the resonance is located below the $K\bar K$ threshold.
Note that in this case the pole from the 2nd quadrant is closer to the physical region of the variable $k$ 
(the physical region of the variable $k$ is indicated by the thick solid lines in 
Figs. \ref{Fig3ab}a,b) and causes the structure in the cross section. 
For $E_{BW}>0$ (Fig. \ref{Fig4ab}b) the resonance 
manifests itself as a bump in the cross section at energies above the $K\bar K$ threshold. 
In this case the pole in the 4th quadrant is located closer to the 
physical region  and is responsible for the structure in the cross section.  

When the coupling to the $K\bar K$ channel is switched on and  
$\bar g_K$ increases, the position of both poles begins moving 
downwards in the (complex) $k$-plane for $E_{BW}>0$ and also for $E_{BW}<0$,
cf. Fig. \ref{Fig3ab}. For very large values of $\bar g_K$ the pole, 
located initially in the 2nd quadrant, moves to the limiting point $k=0$ and the 
second pole moves to $-i\infty$. In this case, according to Eq. (\ref{scaeff}),
the scattering length $a$ goes to infinity and the effective range $r_e$ goes to zero.
In the regime of large coupling, $\bar g_K\gg \bar g_\pi$, 
we have only one pole near the point $k=0$ and the approximate 
expression for elastic scattering amplitude is given again by Eq. (\ref{dlina}). 
This situation corresponds to a molecule-like structure of the near threshold 
resonance \cite{evidence}. Note that most of the available Flatt\'e parametrizations
indicate that the scenario of $\bar g_K\gg \bar g_\pi$ 
is roughly fulfilled for the $f_0(980)$ case, cf. Table 2.
In this context let us emphasize that $\bar g_K\gg \bar g_\pi$
corresponds to a large ratio $R$. The latter quantity can be extracted fairly
reliably from Flatt\'e parametrizations of the experimental invariant mass
distribution as we showed in the last section. 

The corresponding evolution of the cross section with increasing of
the coupling strength can be seen from 
Figs. \ref{Fig4ab}. For $E_{BW}<0$ the initial pure BW resonance evolves into a 
distinct structure that approaches the $K\bar K$ threshold when the channel 
coupling is switched on and gradually increased. The maximum of the cross section 
remains always below the threshold. Clearly, the pole, which is responsible for the 
structure in the cross section for small values of $\bar g_K$ and which was located in the 
2nd quadrant, retains its influence on the resonance shape for all values of $\bar g_K$ 
whereas the second pole (in the 4th quadrant) is not significant.
For the case $E_{BW}>0$ (Fig. \ref{Fig4ab}b) we see again the developement of a 
distinct structure in the cross section which, however, is now a cusp located exactly
at the $K\bar K$ threshold. Also the general situation is different. 
As already said, for small values of $\bar g_K$ the pole located in the 4th quadrant is 
closer to the physical region and is responsible for
the structure in the cross section. With increasing $\bar g_K$ this pole moves 
away from the physical region. 
Simultaneously the pole from the 2nd quadrant moves closer to the physical region 
and takes over the dominant role. Thus we observe here that, with increasing 
$\bar g_K$, one leading pole is substituted by another!

In this context it is interesting to look also at the behaviour 
of the corresponding phase shifts. 
The $S$-matrix corresponding to the Flatt{\'e} amplitude (\ref{nonrel}) is
given by 
\be
S=\eta\exp{(2i\delta(k))}=\frac{E-E_{BW}-i\Gamma_P/2+i\bar g_Kk/2}{E-E_{BW}+i\Gamma_P/2+i\bar g_Kk/2}
\ . 
\label{smatrix}
\ee
Phase shifts $\delta(k)$ for fixed $|E_{BW}|$ and $\Gamma_P$ 
and several values of the coupling constant $\bar g_K$ are presented in Figs. \ref{Fig5ab}. 
The long dashed lines are the results for the uncoupled situation, i.e. for $\bar g_K$ = 0. 
In this case the phase shifts go through $90^{\circ}$ at the nominal BW resonance energy 
$E_{BW}$. 
When the coupling to the $K\bar K$ channel is switched on and increased there is a 
significant difference in the developement of the phases for $ E_{BW} < 0$ and $ E_{BW} > 0$.
For $E_{BW}<0$ (Fig. \ref{Fig5ab}a) the energy where the phase passes through
$90^{\circ}$ moves closer and closer to the threshold and at the same time the 
rise of the phase is getting steeper and steeper. 
Thus, for a strong coupling to the $K\bar K$ channel (i.e. a large $\bar g_K$) 
the behaviour of the phase shift is completely dominated by the occurence of the threshold. 
The parameter values of the 
initial BW resonance ($E_{BW}$ and $\Gamma_P$) play practically no role anymore. 

The results for $E_{BW}>0$ are shown in Fig. \ref{Fig5ab}b. 
As can be seen, for small coupling ($\bar g_K\le0.3$) the phase shifts increase from $0^{\circ}$ to 
$180^{\circ}$ with increasing energy -- whereby a more and more pronounced kink develops at
the $K\bar K$ threshold. 
At a certain value of $\bar g_K$ there is suddenly a jump in the phase and from there
onwards the phase always approaches zero for increasing energy!
This specific behaviour of the phase for $E_{BW}>0$ may be understood by looking at the trajectories 
of the poles, shown in Fig. \ref{Fig3ab}b. 
For $\bar g_K=0$ we have a pair of poles located symmetric with respect to the point $k=0$ and the 
pole in the 4th quadrant is nearest to the physical region of the variable $k$. 
When $\bar g_K$ increases both poles move down in the complex plane. Then, at some stage,
the pole located initially in the 2nd quadrant, crosses the 
real axis of the $k$-plane and reaches the 3rd quadrant. 
(In the specific case shown this occurs for $\bar g_K \approx 0.33$.)  
In the absence of absorption 
(i.e. for $\Gamma_P=0$) this transition corresponds to the mutation of 
a real bound state (in the $K\bar K$ system) to a virtual state. 
This change is reflected also in the $\pi\pi$ system, namely by the mentioned
jump in the phase shift. 
In case of $E_{BW}<0$, the second pole remains in the upper half-plane for all values of $\bar g_K$, 
see Fig. \ref{Fig3ab}a. In this case a bound state exists
for all values of $\bar g_K$ and  the global features of the phase do not change.

\section{The $f_0$ and $a_0$ mesons}


Let us first discuss the $f_0(980)$ resonance. Here the relevant $S$-wave $\pi\pi$ phase
shift in the isospin $I=0$ channel is known experimentally over a wide 
energy range, see, e.g. 
Refs. \cite{Hyams,Protopopescu,Hyams1,bnl} and also the more recent analyses in
Refs. \cite{KLR02,Achasov03}. 
The situation for the $f_0(980)$ resonance differs significantly from 
the ideal situation where there is no background, 
which was discussed in Sections 2 and 3, because the phase 
shift exhibits already a nontrivial behaviour below the $f_0(980)$ region. 
It rises monotonously from the $\pi\pi$ threshold onwards and reaches $90^o$ 
already at an energy around $700$ MeV.
Usually this behaviour is explained via the presence of a broad scalar resonance 
called $\sigma$ or $f_0(400-1200)$ \cite{PDG}  
which is believed to be a pure rescattering effect, see, e.g., \cite{Ulf}. 
This broad resonance provides the background for the $f_0(980)$ 
meson. Because of this large background the $f_0(980)$ meson 
manifests itself as a narrow dip located near the $K\bar K$-threshold 
in some production reactions rather than as a bump \cite{Sassen}. 

Still, because the $\pi\pi$ $S$-wave phase shift is known 
experimentally, at least at a first glance the chances for a determination of the
resonance parameters appear to be much more promising for the channel with 
isospin $I=0$ than for the one with $I=1$ (i.e. for the case of the $a_0$ meson). 
Specifically, one would hope that the knowledge of the $S$-wave phase shift allows 
to fix the scale of the coupling constants $\bar g_{\pi}$ and $\bar g_K$ for 
the $f_0$ case. 
Unfortunately, in practice the situation is much more complicated.
The main problem is, of course, that experimentally only the total phase shift can 
be extracted, and the phase of the background can not be disentangled 
unambiguously from the contribution of the $f_0$ resonance. 
Thus, by making different assumptions about the behaviour of the background one will
always get different solutions for parameters of the $f_0$ resonance. 
The second difficulty is that each of the phase shift analyses is afflicted by 
fairly large error bars. In addition, there are also 
drastic differences between the results of some of those phase shift analyses (compare, 
e.g., Refs. \cite{Hyams,Protopopescu,Hyams1,bnl,KLR02,Achasov03}) for $\pi\pi$ scattering.
These differences preclude the possibility to extract a uniqiue set of parameters for the 
$f_0(980)$ resonance.

To illustrate the two above remarks, we show in Fig. \ref{Fig8} the experimental phases 
taken from Refs. \cite{Hyams,Protopopescu,bnl,KLR02,Achasov03}. 
We also present the $\pi\pi$ phases reconstructed from the
parameters for the $f_0$ meson obtained from different analyses 
\cite{SND,CMD2,KLOE,E791}. 
The latter results exemplify that all parameter sets, and in particular 
those from radiative $\phi$-decays, yield 
very similar phase shifts, despite of the fact that the parameters 
of the $f_0(980)$ resonance differ drastically (see Table 2). 
Thus, they exhibit the typical scaling behaviour that we discussed above. 
Note that for those curves we have added a background phase which
grows smoothly from 70$^o$ to 90$^o$  in the energy region  150
MeV around threshold.
We did this in
order to enable a direct comparison of the calculated results with the energy dependence
of the experimental phase shifts. In all shown cases the same background phase is used. 

The sensitivity of the
inelasticity $\eta$ (the definition of $\eta$ is given by Eq. (\ref{smatrix})) 
to the Flatt\'e parameters is demonstrated in Fig. \ref{Fig9ab} 
for the case of the $f_0(980)$ meson.
Experimental data for the $S$-wave $I=0$ $\pi\pi$ inelasticities are taken from 
Refs. \cite{Hyams,bnl}. Though there are some variations in the results
of different parametrizations it is obvious that todays experimental 
knowledge of the inelasticity in $\pi\pi$ scattering near threshold is not
sufficient for determining the parameters for the $f_0$ resonance. 

Results for the $\pi\pi$ elastic cross section based on those Flatt\'e
parametrizations are shown in Fig. \ref{Fig12}. We can see again that most
of those parametrizations yield rather similar predictions. 

Now let us discuss shortly the $a_0$ meson. As is seen from Tables 1 and 2 the strength of 
the coupling to the $K\bar K$-channel, $\bar g_K$, is much weaker for the $a_0$ meson, 
than for the $f_0$ meson whereas the coupling to the $\pi\pi$ or $\pi\eta$ channels
is comparable. Accordingly, the ratio $R$ is, in general, significantly smaller for
the $a_0$ case. As a consequence, for the $a_0$ meson the magnitude of the 
effective range is much larger and both singularities of the scattering amplitude 
influence the differential cross section near the $K\bar K$ threshold. 
These aspects were already discussed in our previous paper \cite{evidence}. 
  
The behaviour of the $\pi\eta$ $S$-wave phase shift 
is presented in Fig. \ref{Fig11} for the Flatt\'e parametrizations
given in Refs. \cite{Teige,Bugg,KLOEa,AchKi}. Also here one sees that 
there is basically no difference in the prediction for the phase shifts 
despite the fact that all  sets of parameters are quite different. 
However, the situation is somewhat different for the $\pi\eta$ inelasticity 
$\eta$ which is shown in Fig. \ref{etapieta} and also for the $\pi
\eta$ total cross section (cf. Fig. \ref{xspieta}). Here there are stronger variations
between the results produced by the different parametrizations. 
Thus,  the scaling behavior is not so pronounced for the 
case of the $a_0$ meson, which could be already guessed from the
smaller ratio $R$ (cf. discussion in section
\ref{distrib}). Consequently, for the $a_0$ resonance there could be a better
chance to determine all three Flatt{\'e} parameters as compared to the
$f_0$ meson.

These examples show that the determination of the ``true'' values for the parameters of 
the $a_0$ as well as the $f_0$ meson is a rather challenging problem. It requires, 
first of all, a very significant improvement in the accuracy of the experimental 
data.

\section{Conclusions}

In the present paper we studied properties of the Flatt{\'e} amplitude (\ref{nonrel}),
which is usually employed to describe differential mass distributions resulting from
$S$-wave resonance like structures, located near a threshold. Specifically, such
Flatt{\'e} and Flatt{\'e}-like distributions are often used to describe and
derive properties of the scalar mesons $a_0(980)$ and $f_0(980)$. But there are 
also some other examples of hadronic resonances, located near thresholds, where it 
is reasonable to represent them in terms of the Flatt{\'e} distribution.

The Flatt{\'e} parametrization of the amplitude includes three free parameters 
which should be determined from the experimental mass spectrum. These are
the nominal mass of the resonance, $m_R$, the inelastic width at threshold, $\Gamma_P$, 
(or the coupling constant for the channel of the light particles, $\bar g_P$), where
$P$ stands for $\pi\pi$ or $\pi\eta$, and the coupling constant $\bar g_K$ for 
the channel of the heavy particles. 

We showed that the mass spectrum near threshold is not sensitive to all the  
parameters ($E_{BW}$, $\bar g_P$, $\bar g_K$) but rather only to the two 
dimensionless ratios $R=\bar g_K/\bar g_P$ and $\alpha= 2E_{BW}/\Gamma_P$. 
Those are the two parameters that 
determine also the scattering length in the channel with the heavy particles. 
This difficulty of fixing all three Flatt{\'e}  parameters concerns the standard
Flatt\'e distribution but also relativistic extensions like the one 
proposed by Achasov and is clearly reflected in the parameter values for the
$a_0$(980) and $f_0$(980) mesons that can be found in the literature (cf. 
Tables 1 and 2). The results of those fits to the data exemplify 
that there is a large uncertainty in the absolute values of the coupling constants, 
whereas the ratios $R$ and $\alpha$ can be extracted from experiments with
much better accuracy.

In principle, only 
the information about the absolute values of the coupling constants and the 
nominal resonance mass opens the possibility to calculate the $K\bar K$ 
effective range parameters and to reconstruct the position 
of the poles of the scattering amplitude in the complex $k$-plane. 
As was stressed in Ref. \cite{evidence} and  
in line with the suggestion made by  Morgan \cite{morgan}, the knowledge
of the position of the poles of the scattering amplitude
allows to draw conclusions on the nature of the resonance, i.e. to 
clarify, whether it corresponds to an elementary object made from quarks or  
whether it is a compound state like a $K\bar K$ molecule. 
However, the ratio $R$ of the coupling constants is an interesting
quantity too. For example, a large $R$, i.e. a large coupling to the 
$K\bar K$ channel is a strong indication for a molecular-like
structure of the near-threshold resonance. 
Fortunately, as we have shown, this ratio can be determined from a Flatt{\'e} 
parametrization of the mass distributions with significantly better 
reliability than the absolute values of the coupling constants.

\acknowledgement{
We would like to thank Yu. Kalashnikova for instructing discussions
and suggestions. 
This work was supported by the DFG-RFBR grant no.
02-02-04001 (436 RUS 113/652). A.E.K. acknowledges  also the partial support
by the grant RFBR 02-02-16465.
} 

\vspace*{-0.5cm}

\section{Appendix}

The elastic scattering amplitude for  the channel with the light
particles for the relativistic Flatt\'e
distribution has the form
\be
f_{el}=-\frac{1}{q}\, \frac{\Gamma_P m_R}{m^2-m_R^2+im_R(\Gamma_P+\bar
 g_k\frac{k}{2})} \ .
\ee
In the nonrelativistic limit it can be rewritten as
\be
f_{el}=-\frac{1}{2q}\frac{\Gamma_P}{E-E_{BW}+i\frac{\Gamma_P}{2}+i\bar g_K\frac{k}{2}} \ .
\label{NR}
\ee
where we used the notation introduced in Section \ref{distrib}.

Let us demonstrate in this Appendix that the relativistic Flatt\'e-like distribution
introduced by Achasov \cite{AchKi}, i.e. the expression
\be
f_{el}=-\frac{1}{q}\, \frac{\Gamma_P m_R}{m^2-m_R^2+\sum_{ab}
[\rm\Pi^{ab}(m^2)-Re\Pi^{ab}(m_R^2)]},
\label{ach}
\ee
where $\rm \Pi$ takes into account 
the so called finite width corrections to the self-energy loop of the
resonance with nominal mass $m_R$ from the two-particle intermediate 
states $ab$ ($\pi\pi$ or $\pi\eta$ and $K\bar K$), 
reduces to the form given in Eq. (\ref{NR}) in the nonrelativistic limit.
In the region $m\ge m_a+m_b$ the expression for $\rm\Pi^{ab}$ is 
\begin{eqnarray}
\label{polarisator}
&&\Pi^{ab}(m^2)=\frac{g^2_{ab}}{16\pi}\left[\frac{m_+m_-}{\pi
m^2}\ln \frac{m_b}{m_a}+\right.\nonumber\\
&&\left.+\rho_{ab}\left(i+\frac{1}{\pi}\ln\frac{\sqrt{m^2-m_-^2}-
\sqrt{m^2-m_+^2}}{\sqrt{m^2-m_-^2}+\sqrt{m^2-m_+^2}}\right)\right],
\end{eqnarray}
where $m_{\pm}=m_a\pm m_b$ with $m_a\ge m_b$ and \\
$\rho_{ab}(m)=\sqrt{(1-\frac{m_+^2}{m^2})(1-\frac{m_-^2}{m^2})}\,$.
In the nonrelativistic limit the expressions for $\Delta{\rm  \Pi}^{ab}/4m_K= \left [ {\rm Re\: \Pi}^{ab}(m^2)-
 {\rm Re\: \Pi}^{ab}(m_R^2)\right ]/4m_K$ take the form
\begin{eqnarray}
\nonumber
&&\frac{ \Delta\: {\rm \Pi}^{K\bar K}}{4m_K}
\approx\frac{\bar g_K}{\pi}(E-E_R)\\ 
&&\frac{ \Delta\: {\rm \Pi}^{P}}{4m_K}=\frac{\bar
  g_{P}}{4\pi}(E-E_R)\: C_P
\label{loop}
\end{eqnarray}
where $P$ denotes $\pi\pi$ or $\pi\eta$ loops and
\begin{eqnarray}
\nonumber
&&C_P=\left \{\frac{m_+m_-}{2m_k^2}\ln\frac{m_b}{m_a}\right.\\
&&\hspace*{-1cm}\left.
-A_-A_+\left[(B_-+B_+)\ln\frac{A_--A_+}{A_-+A_+}+\frac{A_-A_+(B_--B_+)}{A_-^2-A_+^2} 
\right ]\right \},
\end{eqnarray}
\begin{eqnarray}
 A_{\pm}=\sqrt{1-\frac{(m_a\pm m_b)^2}{4m_K^2}},\:
B_{\pm}=\frac{(m_a\pm m_b)^2}{4m_K^2-(m_a\pm m_b)^2}.
\end{eqnarray}
Note that the expression for  $C_{\pi\pi}$  can be easily simplified to 
\be
C_{\pi\pi}=2\left (1-\frac{m_{\pi}^2}{m_K^2}\ln\frac{m_{\pi}}{2m_K}\right
)+O\left (\frac{m_{\pi}^2}{m_K^2}\right )\approx 2.32,
\ee
whereas the evalution of $C_{\pi\eta}$ gives the value 2.81.

Thus, using the nonrelativistic expansion of the loop corrections
(Eqs.\ref{loop})  Eq.(\ref{ach}) can be reduced to the standard Flatt\'e form.
For example, for the $f_0$ meson, one gets
\bea
\nonumber
\hspace*{-0.5cm}&&f_{\pi\pi}=-\!\frac{1}{2q}\times\\
\nonumber
&&\hspace*{-0.5cm} \frac{\Gamma_P}{(E\!-\!E_R)(1\!-\!\frac{\bar g_K}{\pi}\!-\!\frac{\bar g_{\pi}}{4\pi}C_{\pi\pi})
\!+\!\frac{\bar g_K}{2}\sqrt{m|E_R|}\Theta(\!-E_R)
+i\frac{\Gamma_P}{2}+i\bar g_K\frac{k}{2}}\\
&=&-\frac{1}{2q}\, \frac{\tilde \Gamma_P}{E\!-E_{BW}
\!+\!i\frac{\tilde \Gamma_P}{2}\!+\!i\tilde  g_K\frac{k}{2}},
\label{achflat}
\eea
where
\bea
\nonumber
&& \tilde  g_{\pi}=
\frac{\bar g_{\pi}}{1-\dis\frac{\bar g_K}{\pi}-\frac{\bar g_{\pi}}{4\pi}
  C_{\pi\pi}}, \ 
\tilde  g_{K}=
\frac{\bar g_{K}}{1-\dis \frac{\bar g_K}{\pi}-\dis\frac{\bar
    g_{\pi}}{4\pi}  C_{\pi\pi}},\\
&&\tilde \Gamma_P=\tilde  g_{\pi}q,\\
&&E_{BW}=\!E_R\!-\!\dis\frac{\tilde
  g_K}{2}\sqrt{m|E_R|}\Theta(-E_R), \\ 
&&\alpha=\frac{2E_{BW}}{\tilde \Gamma_P}\ . 
\label{achparam}
\eea
Note that the term $\dis\frac{\bar g_K}{2}\sqrt{m|E_R|}\Theta(-E_R)$
in the denominator results from the piece $\rm Re\Pi(m_R^2)$ in the
case of a negative $E_R$. We want to mention also 
that for some values of the parameters $\bar g_K$ and $\bar g_{\pi}$ 
the expression ${1-\dis\frac{\bar g_K}{\pi}-\frac{\bar g_{\pi}}{4\pi}
C_{\pi\pi}}$ can change the sign. In this case the effective range
in Eq. (\ref{scaeff}) also changes its sign and becomes positive.
Then the relativistic distribution introduced by Achasov (Eq. (\ref{ach}))
no longer can be reduced to the standard nonrelativistc form (\ref{NR})
but it will go over into what one might call an Anti-Flatt\'e 
distribution, i.e. into a form that is identical to Eq. (\ref{NR})
but where the sign of the real part in the denominator is reversed. 


\newpage 
 
\begin{figure}[h]
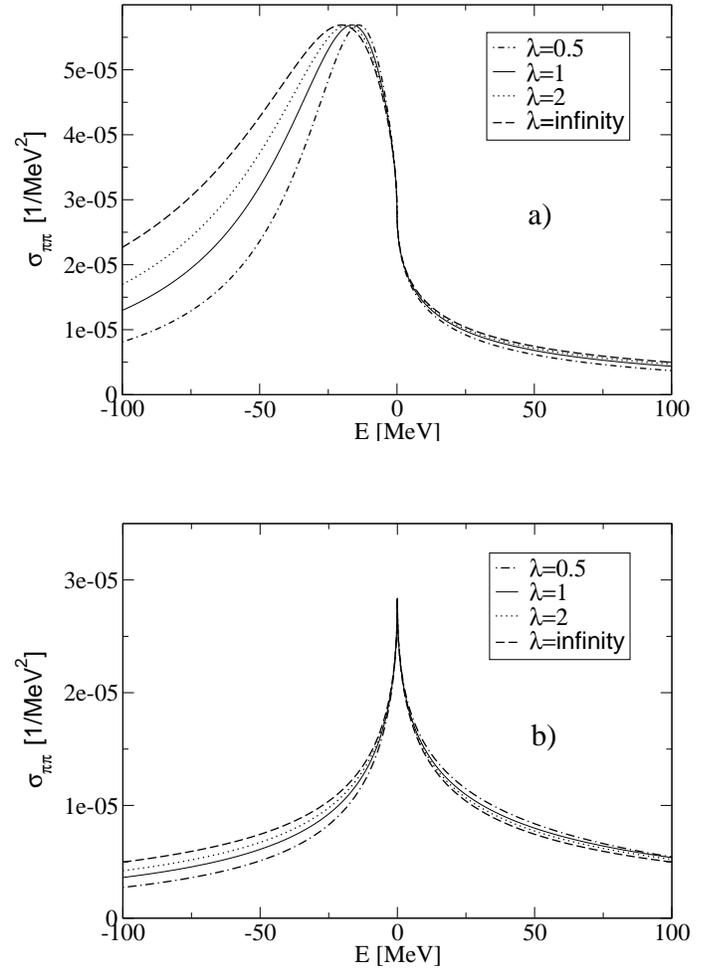

\begin{center}
\epsfig{file=xsscaling.eps,width=9cm}

\vspace*{1.cm}

\epsfig{file=xsscalingEfplus163.eps,width=9cm}
\caption{$\pi\pi$ cross section for different scaling parameters
$\lambda$.  
The results are based on the Flatt\'e parameters $R=4.76$, $\bar g_\pi$ = 0.317, 
$\bar g_K$ = 1.51, and $|E_{BW}|$ = 163.7 MeV of Ref. \protect\cite{CMD2}. 
a) $E_{BW}<0$ , b) $E_{BW}>0$. 
} 
\label{Fig1ab}
\end{center}
\end{figure}

\begin{figure}[h]
\begin{center}
\vspace*{0.5cm}
\epsfig{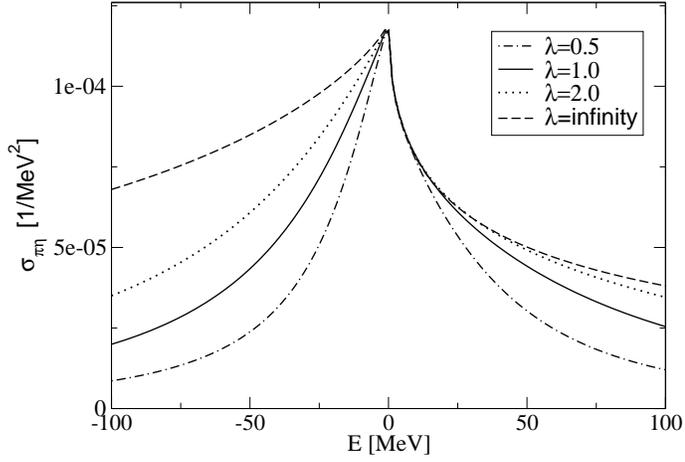}
\caption{$\pi\eta$ cross section for different scaling parameters
$\lambda$.  
The results are based on the Flatt\'e parameters $R=1.14$, $\bar g_\pi$ = 0.454, 
$\bar g_K$ = 0.506, and $E_{BW}$ = 7.6 MeV of Ref. \protect\cite{Bugg}. 
} 
\label{Fig1a0}
\end{center}
\end{figure}

\begin{figure}[h]
\begin{center}
\epsfig{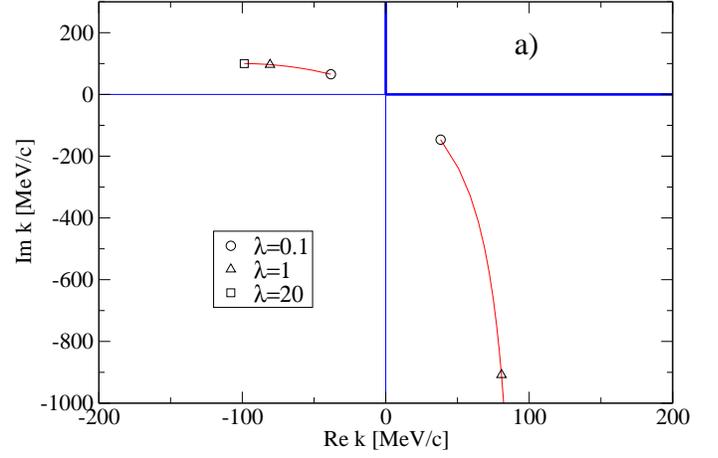}

\vspace*{1.3cm}

\epsfig{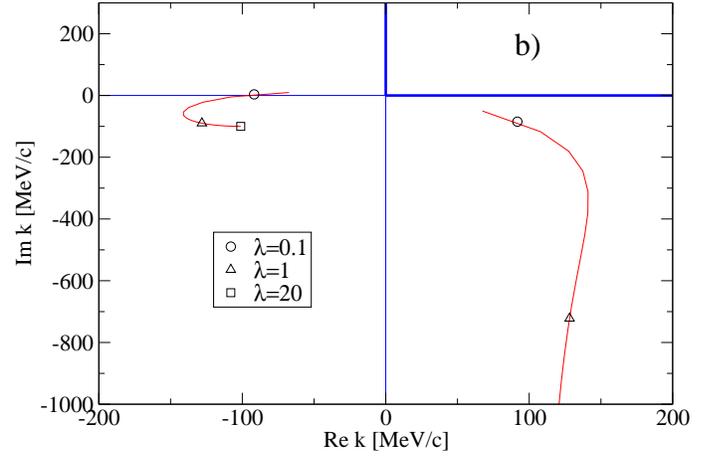}
\caption{Trajectories of the poles in the complex $k$-plane, where $k$ is the
cms momentum in the $K\bar K$ system. 
The same Flatt\'e parameters as in Fig. \ref{Fig1ab} are used.
The scaling parameters $\lambda$ is varied. 
a) $E_{BW}<0$ , b) $E_{BW}>0$.
The physical region of the variable $k$ is indicated by the thick solid lines.
}
\label{Fig2ab}
\end{center}
 \end{figure}

\begin{figure}[h]
\begin{center}
\epsfig{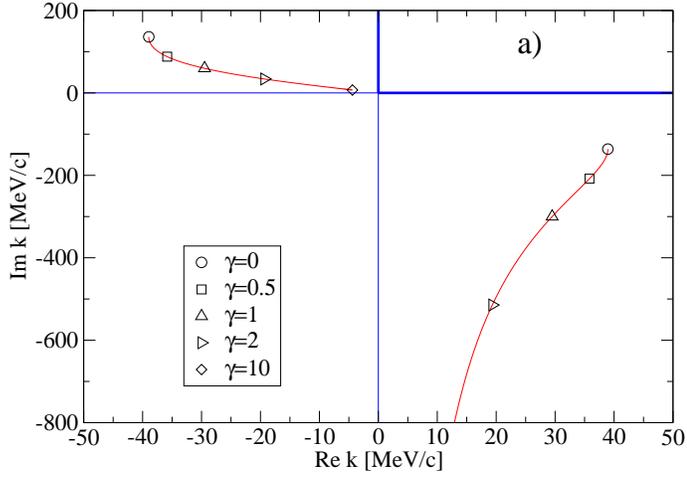}

\vspace*{1.3cm}
\epsfig{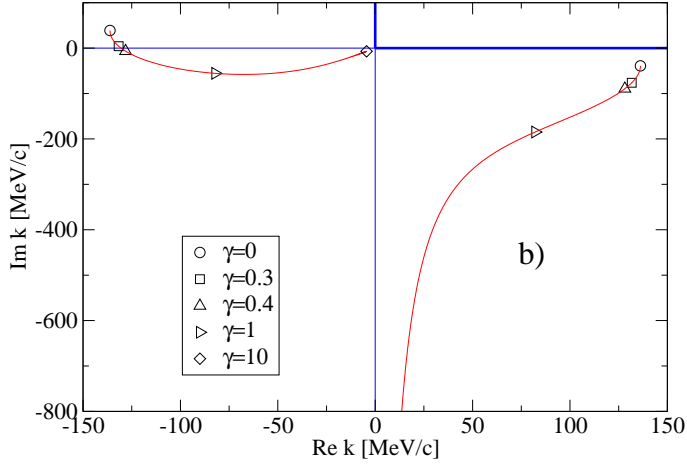}
\caption{Trajectories of the poles in the complex $k$-plane, where $k$ is the
cms momentum in the $K\bar K$ system.
The results are based on the Flatt\'e parameters $\bar g_\pi$ = 0.09, 
$\bar g_K$ = 0.97, and $|E_{BW}|$ = 34.3 MeV of Ref. \protect\cite{OPAL}.
The coupling strength to the $K\bar K$ channel is varied by $\gamma \times \bar g_K$.
a) $E_{BW}<0$ , b) $E_{BW}>0$. 
The physical region of the variable $k$ is indicated by the thick solid lines.
}
\label{Fig3ab}
\end{center}
\end{figure}

\begin{figure}[h]
\begin{center}
\epsfig{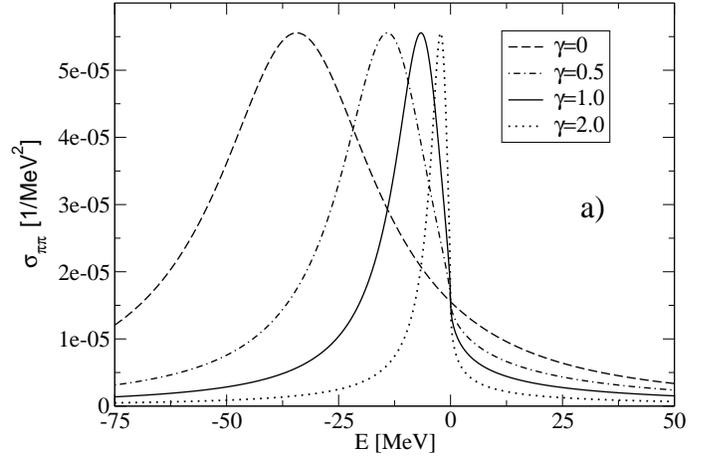}

\vspace*{1.5cm}

\epsfig{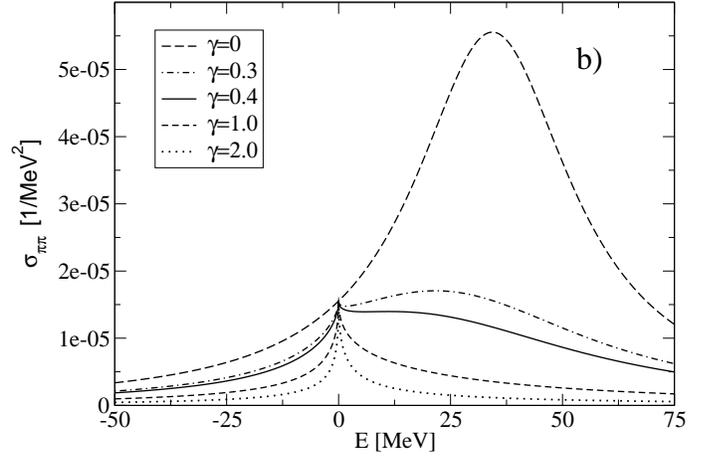}
\caption{Results for the $\pi\pi$ cross section. 
The same Flatt\'e parameters as in Fig. \ref{Fig3ab} are used.
The coupling strength to the $K\bar K$ channel is varied by 
$\gamma \times \bar g_K$.  
a) $E_{BW}<0$, b) $E_{BW}>0$.
} 
\label{Fig4ab}
\end{center}
\end{figure}

\begin{figure}[h]
\begin{center}
\epsfig{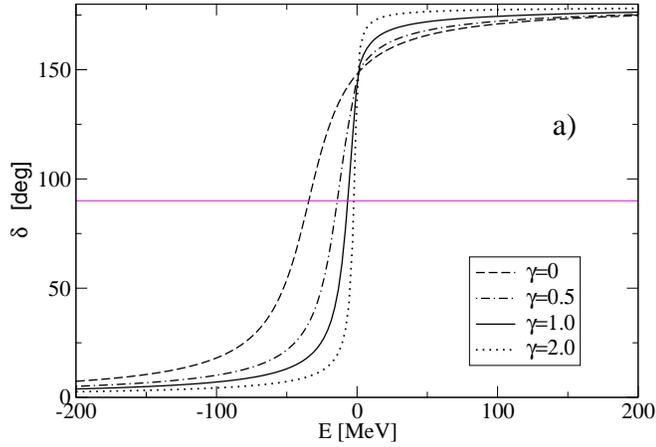}


\vspace*{1.5cm}

\epsfig{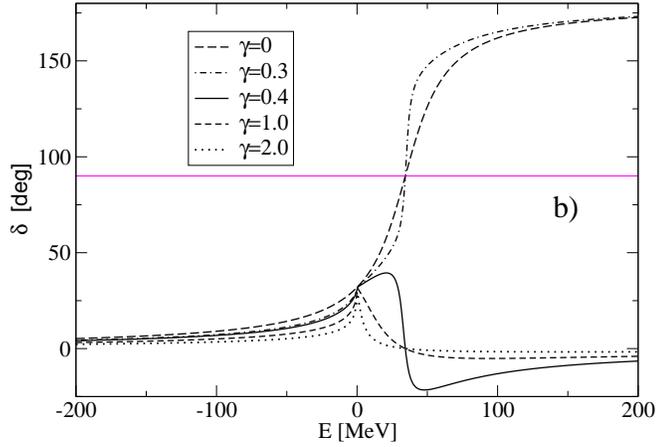}
\caption{$\pi\pi$ phase shift $\delta$ in the 
$J=0$, $I=0$ partial wave. 
The same Flatt\'e parameters as in Fig. \ref{Fig3ab} are used.
The coupling strength to the $K\bar K$ channel is varied by 
$\gamma \times \bar g_K$.
a) $E_{BW}<0$, b) $E_{BW}>0$ } 
\label{Fig5ab}
\end{center}
\end{figure}

\vskip 6.5cm

\begin{figure}[h]
\begin{center}
\epsfig{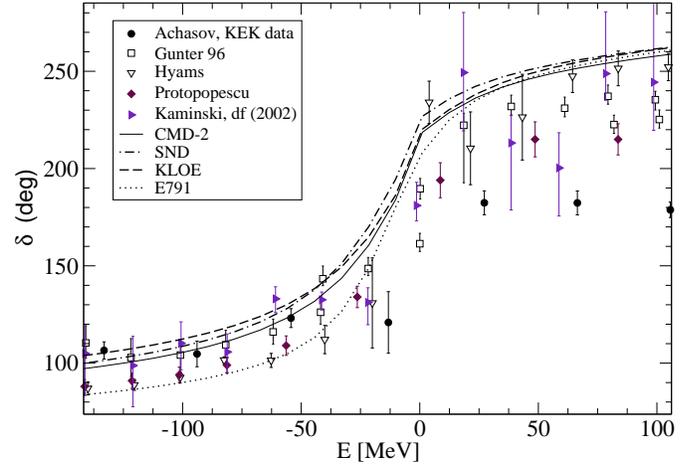}
\caption{$\pi\pi$ phase shift $\delta$ in the
$J=0$, $I=0$ partial wave. 
The curves are results based on Flatt\'e 
distributions taken from Refs. \cite{SND,CMD2,KLOE,E791}.
The symbols show results from various phase shift analyses 
taken from Refs. \cite{Hyams,Protopopescu,bnl,KLR02,Achasov03}. 
} 
\label{Fig8}
\end{center}
\end{figure}

\begin{figure}[h]
\begin{center}
\vspace*{1.0cm}
\hspace*{0.5cm}
\epsfig{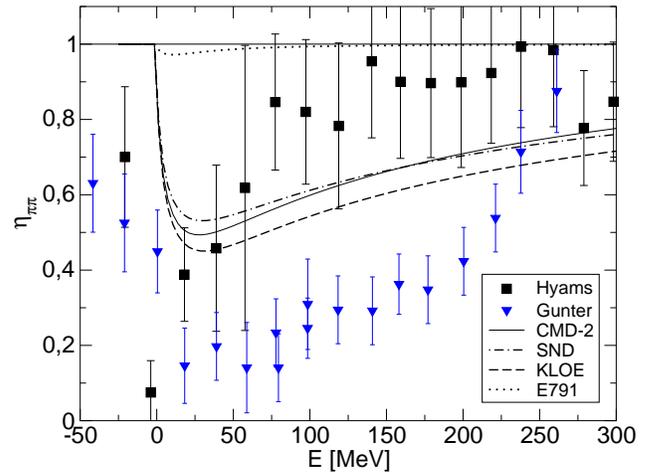}
\caption{Inelasticity in the $\pi\pi$ $J=0$, $I=0$ 
partial wave. The curves are results based on Flatt\'e
parameters taken from Refs. \cite{SND,CMD2,KLOE,E791}.
The symbols show results from phase shift analyses
taken from Refs. \cite{Hyams,bnl}.
} 
\label{Fig9ab}
\end{center}
\end{figure}

\begin{figure}[h]
\begin{center}
\vspace*{0.4cm}
\epsfig{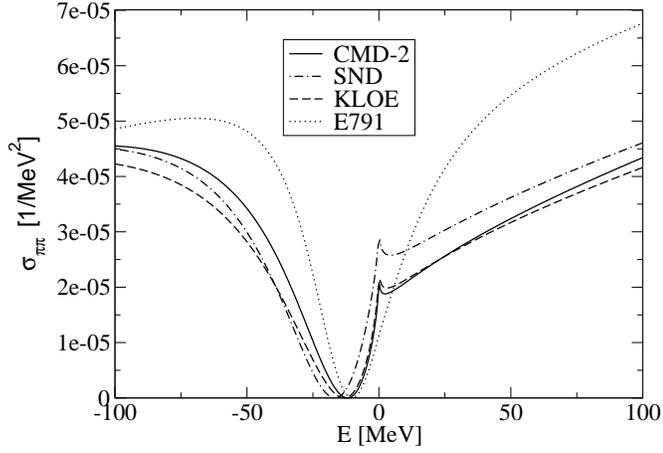}
\caption{Results for the $\pi\pi$ cross section. 
The curves are results based on Flatt\'e 
distributions taken from the Refs. \cite{SND,CMD2,KLOE,E791}.} 
\label{Fig12}
\end{center}
\end{figure}

\begin{figure}[h]
\begin{center}
\vspace*{1.0cm}
\epsfig{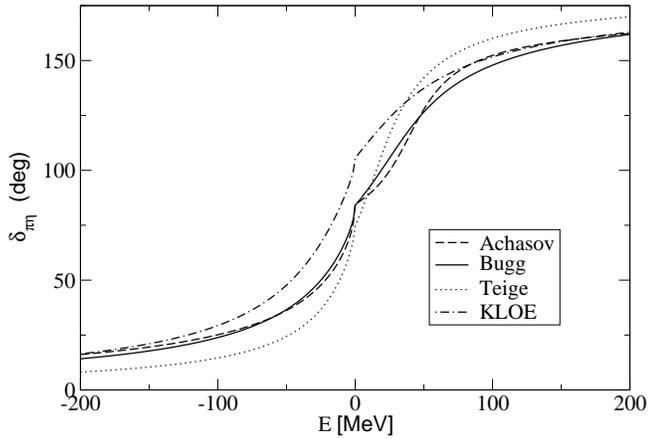}
\caption{$\pi\eta$ phase shift $\delta$ 
in the $J=0$, $I=1$ partial wave. 
The curves are results based on Flatt\'e 
distributions taken from Refs. \cite{Teige,Bugg,KLOEa,AchKi}.
} 
\label{Fig11}
\end{center}
\end{figure}

\begin{figure}[h]
\begin{center}
\vspace*{0.4cm}
\hspace*{0.5cm}
\epsfig{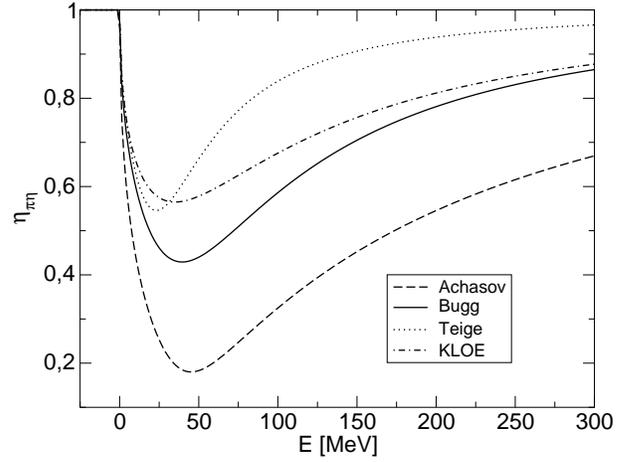}
\caption{Inelasticity in the 
$\pi\eta$ $J=0$, $I=1$ partial wave. 
The curves are results based on Flatt\'e
parameters taken from Refs. \cite{Teige,Bugg,KLOEa,AchKi}.
} 
\label{etapieta}
\end{center}
\end{figure}

\begin{figure}[h]
\begin{center}
\vspace*{0.4cm}
\hspace*{0.5cm}
\epsfig{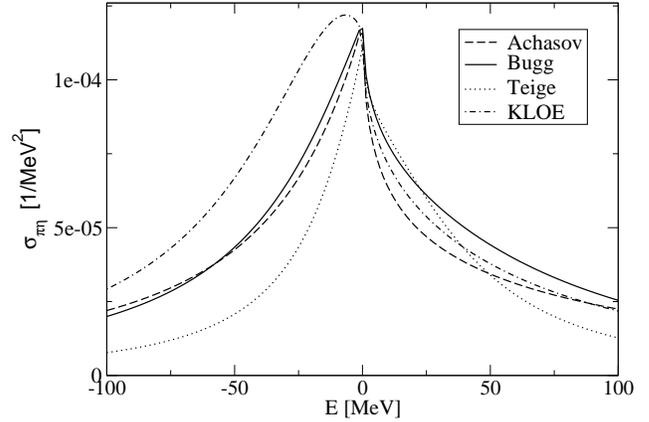}
\caption{Total cross section in the 
$\pi\eta$ $J=0$, $I=1$ partial wave. 
The curves are results based on Flatt\'e
parameters taken from Refs. \cite{Teige,Bugg,KLOEa,AchKi}.
} 
\label{xspieta}
\end{center}
\end{figure}


\begin{thebibliography}{99}

\bibitem{Buggrep} D.~V.~Bugg,
Phys.\ Rept.\  {\bf 397}, 257 (2004).

\bibitem{Klempt} E. Klempt, hep-ph/0404270. 

\bibitem{Torn} F.E. Close and N.A. Tornqvist, 
J. Phys. {\bf G 28}, R249 (2002); 
C. Amsler and N.A. Tornqvist, 
Phys. Rept. {\bf 389}, 61 (2004). 

\bibitem{Oller} J.A. Oller, E. Oset , and A. Ramos,
Prog. Part. Nucl. Phys. {\bf 45}, 157 (2000). 


\bibitem{Bev} E. van Beveren and G. Rupp, hep-ph/0406242. 

\bibitem{Ani} V.V. Anisovich, Phys.\ Usp.\  {\bf 47}, 45 (2004)
[Usp.\ Fiz.\ Nauk {\bf 47}, 49 (2004)]; arXiv:hep-ph/0208123.


\bibitem{Flatte} S. Flatt{\'e}, Phys. Lett. {\bf 63B}, 224 (1976).


\bibitem{Teige} S. Teige et al., Phys. Rev. D {\bf 59}, 012001 (1999).

\bibitem{Bugg} D.V. Bugg, V.V. Anisovich, A. Sarantsev, and B.S. Zou, 
Phys. Rev. D {\bf 50}, 4412 (1994).

\bibitem{Amsler} A. Abele et al., Phys. Rev. D {\bf 57}, 3860 (1998).

\bibitem{SNDa} M.N. Achasov et al., Phys. Lett. {\bf B479}, 53 (2000).

\bibitem{KLOEa} A. Aloisio et al., Phys. Lett. {\bf B536}, 209 (2002). 

\bibitem{AchKi} N.N. Achasov and A.N. Kisilev,
Phys. Rev. D {\bf 68}, 014006 (2003). 


\bibitem{SND} M.N. Achasov et al., Phys. Lett. {\bf B485}, 349 (2002).

\bibitem{CMD2} R.R. Akhmetshin et al., Phys. Lett. {\bf B462}, 380 (1999).

\bibitem{KLOE} A. Aloisio et al., Phys. Lett. {\bf B537}, 21 (2002); 
A. Antonelli, hep-ex/0209069 (2002).

\bibitem{E791}
E.~M.~Aitala {\it et al.}  [E791 Collaboration],
Phys.\ Rev.\ Lett.\  {\bf 86}, 765 (2001).

\bibitem{WA102}
D.~Barberis {\it et al.}  [WA102 Collaboration],
Phys.\ Lett.\ B {\bf 462}, 462 (1999).

\bibitem{OPAL}
K.~Ackerstaff {\it et al.}  [OPAL Collaboration],
Eur.\ Phys.\ J.\ C {\bf 4}, 19 (1998).


\bibitem{AchGu} N.N. Achasov and V.V. Gubin, Phys. Rev. D {\bf 63}, 094001 (2001).

\bibitem{xxx} see, e.g., R.G. Newton, {\it Scattering Theory of Waves and
Particles}, (Springer-Verlag, New York, 1982). 

\bibitem{yyy}  M.M. Nagels, T.A. Rijken, and J.J. de Swart, 
Phys. Rev. D {\bf 20}, 1633 (1979); 
B. Holzenkamp, K. Holinde, and J. Speth, Nucl. Phys. {\bf A500}, 485 (1989).

\bibitem{Kerbikov} B. Kerbikov, hep-ph/0402022. 

\bibitem{evidence} V. Baru, J. Haidenbauer, C. Hanhart, Yu.S. Kalashnikova,
and A. Kudryavtsev, Phys.\ Lett.\ B {\bf 586}, 53 (2004);
Yu.S. Kalashnikova, arXiv:hep-ph/0311302.

\bibitem{morgan} D. Morgan, Nucl. Phys. {\bf A543}, 632 (1992).

\bibitem{Hyams}
B.~Hyams {\it et al.},
Nucl.\ Phys.\ B {\bf 64}, 134 (1973)
[AIP Conf.\ Proc.\  {\bf 13}, 206 (1973)].

\bibitem{Protopopescu}
S.~D.~Protopopescu {\it et al.},
Phys.\ Rev.\ D {\bf 7}, 1279 (1973).

\bibitem{Hyams1} B. Hyams et al., Nucl. Phys. {\bf B100}, 205 (1975).

\bibitem{bnl} 
J.~Gunter {\it et al.}  [E852 Collaboration],  
arXiv:hep-ex/9609010.

\bibitem{KLR02}
R.~Kaminski, L.~Lesniak and K.~Rybicki,
Eur.\ Phys.\ J.\ directC {\bf 4}, 4 (2002)
[arXiv:hep-ph/0109268].

\bibitem{Achasov03}
N.~N.~Achasov and G.~N.~Shestakov,
Phys.\ Rev.\ D {\bf 67}, 114018 (2003).

\bibitem{PDG} K. Hagiwara et al., Phys. Rev. D {\bf 66}, 010001 (2002).

\bibitem{Ulf} U.-G. Mei\ss ner, Comm. Nucl. Part. Phys. 
{\bf 20}, 119 (1991). 

\bibitem{Sassen} F.P. Sassen, S. Krewald, and J. Speth, 
Eur. Phys. J. A {\bf 18}, 197 (2003).  

\end{thebibliography}
\end{document}